\documentclass{iau}
\usepackage{graphicx}
\DeclareGraphicsExtensions{.pdf,.png,.jpg}
\usepackage{floatrow}
\usepackage[singlelinecheck=false]{caption}

\title[Euclid space mission] 
{Euclid space mission: a cosmological challenge for the next 15 years}

\author[Roberto Scaramella et al.] 
{R. Scaramella$^1$, Y. Mellier, J. Amiaux, C. Burigana, C.S. Carvalho, J.C. Cuillandre, A. da Silva, J. Dinis,  A. Derosa, E. Maiorano,  P. Franzetti, B. Garilli, M. Maris,   M. Meneghetti,  I. Tereno, S. Wachter, L. Amendola, M. Cropper, V. Cardone, R. Massey, S. Niemi, H. Hoekstra, T. Kitching, L. Miller, T. Schrabback,  E. Semboloni,  A. Taylor, M. Viola,  T. Maciaszek, A. Ealet, L. Guzzo, K. Jahnke, W. Percival, F. Pasian, M. Sauvage \and the Euclid Collaboration}
\affiliation{$^1$Osservatorio di Roma \\ email: {\tt kosmobob@oa-roma.inaf.it} 
}

\pubyear{2014}
\volume{306}  
\pagerange{??--??}
\jname{Statistical Challenges in 21st Century Cosmology}
\editors{A.F. Heavens, J.-L. Starck \& A. Krone-Martins, eds.}
\begin{document}

\maketitle
\begin{abstract}
Euclid is the next ESA mission devoted to cosmology. It aims at observing most of the extragalactic sky, studying both gravitational lensing and clustering over $\sim$15,000 square degrees. The mission is expected to be launched in year 2020 and to last six years. The sheer amount of data of different kinds, the variety of (un)known systematic effects and the complexity of measures require efforts both in sophisticated simulations and techniques of data analysis. 
We  review the mission main characteristics, some aspects of the the survey  and highlight some of the areas of interest to this meeting.
\end{abstract}
\firstsection 
%
\section{Overview}
 \noindent {\underline{\it Science}}.
Euclid is an ESA Cosmic Vision mission (\cite{redbook}) mainly devoted to the study of cosmology, namely precise measurements of the expansion factor $a(t)$ and of the growth of density perturbations, $\delta \rho/\rho$, whose power spectrum is $P(k)$. This will be achieved via two main probes: (i)  gravitational lensing [WL] via exquisite shape measures + photoz of over a billion of galaxies  and (ii) the clustering of galaxies via accurate redshift mostly obtained by emission lines, measured through a slitless spectrograph.

An important aspect of Euclid is the contemporary use of additional cosmological probes (to WL and clustering add clusters of galaxies -- counts, density profiles and mass function --, the Integrated Sachs-Wolfe, and possibly high-z SNe) and their overall  complementarity.  Moreover, it is  possible to study not only the geometry but also the detailed evolution of density perturbations: the information on the growth factor (usually parametrised as $d \ln \delta / d \ln a\propto \Omega_m^\gamma$, where for dark matter dominated models is $\gamma \sim 4/7$) allows one to better differentiate among competing classes of models, such as dark energy vs. modified gravity (see Fig. 1, and \cite{theory}).

\begin{figure}[h!]
\begin{center}
 \includegraphics[width=13.5cm]{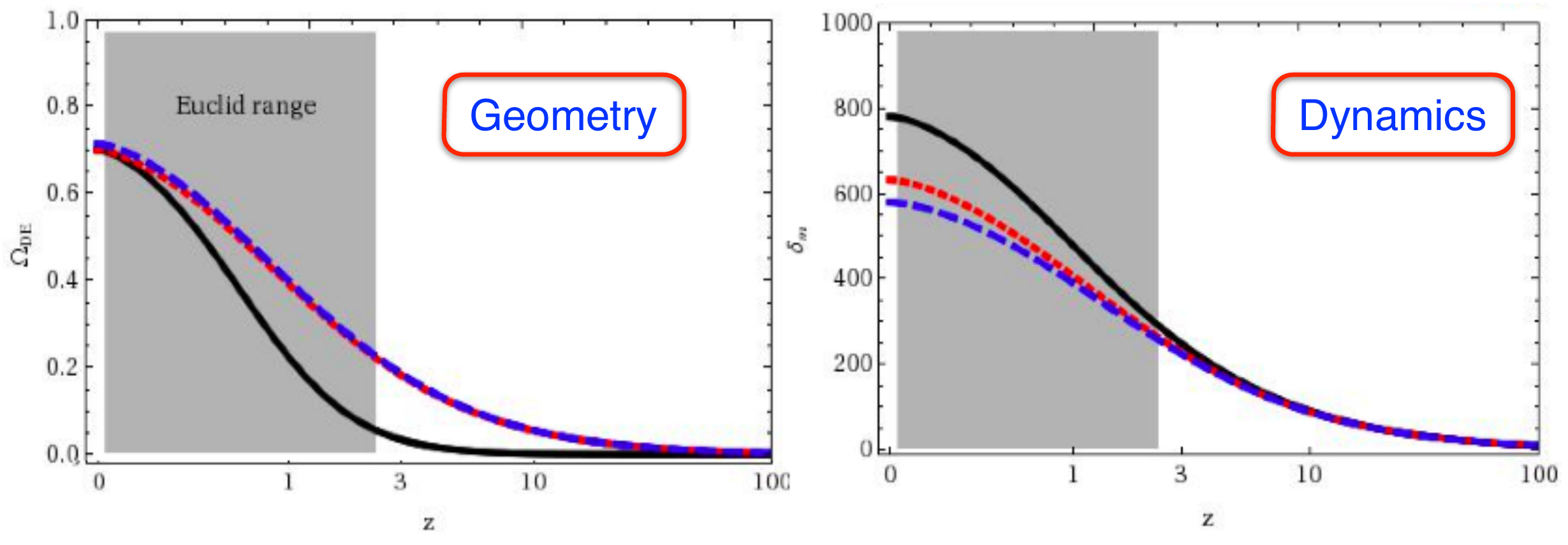} 
  \caption{Effect of dark energy on the evolution of the Universe. Left: Fraction of the density of the Universe in the form of dark energy as a function of redshift, z, for a model with a cosmological constant ($w=-1$, black solid line), and two models with identical background expansion: quintessence with a different equation of state (blue dashed line), and a modified gravity model -DGP- (red dotted line, with $w_0=-0.77$ and $w_a=0.29$, with $w=w_0+w_a(1+z)/z$). In all cases, dark energy becomes dominant in the low redshift Universe era probed by Euclid. Right: Growth factor of cosmic structures for the same three models. Only by adding at low redshifts  the measure of the growth of structure (right panel) to measure of the geometry (left panel)   a modification of dark energy can be distinguished from that of gravity because of the different predicted growth rate. Weak lensing and clustering measure both effects.}
   \label{fig1}
\end{center}
\vspace*{-0.50 cm}
\end{figure}

Moreover, both the variety of data and the large sky coverage ($\sim 15,000$ sq. deg. in the wide survey) will yield a wealth of information which will be crucial for several areas in astronomy, yielding a large legacy value to Euclid databases. Also the deep fields ($\sim 40$ sq. deg. observed repeaditly to reach two magnitudes fainter than the wide survey) will be unique for these purposes (\cite{redbook}).

\begin{figure}[b]
\begin{center}
 \includegraphics[width=10cm]{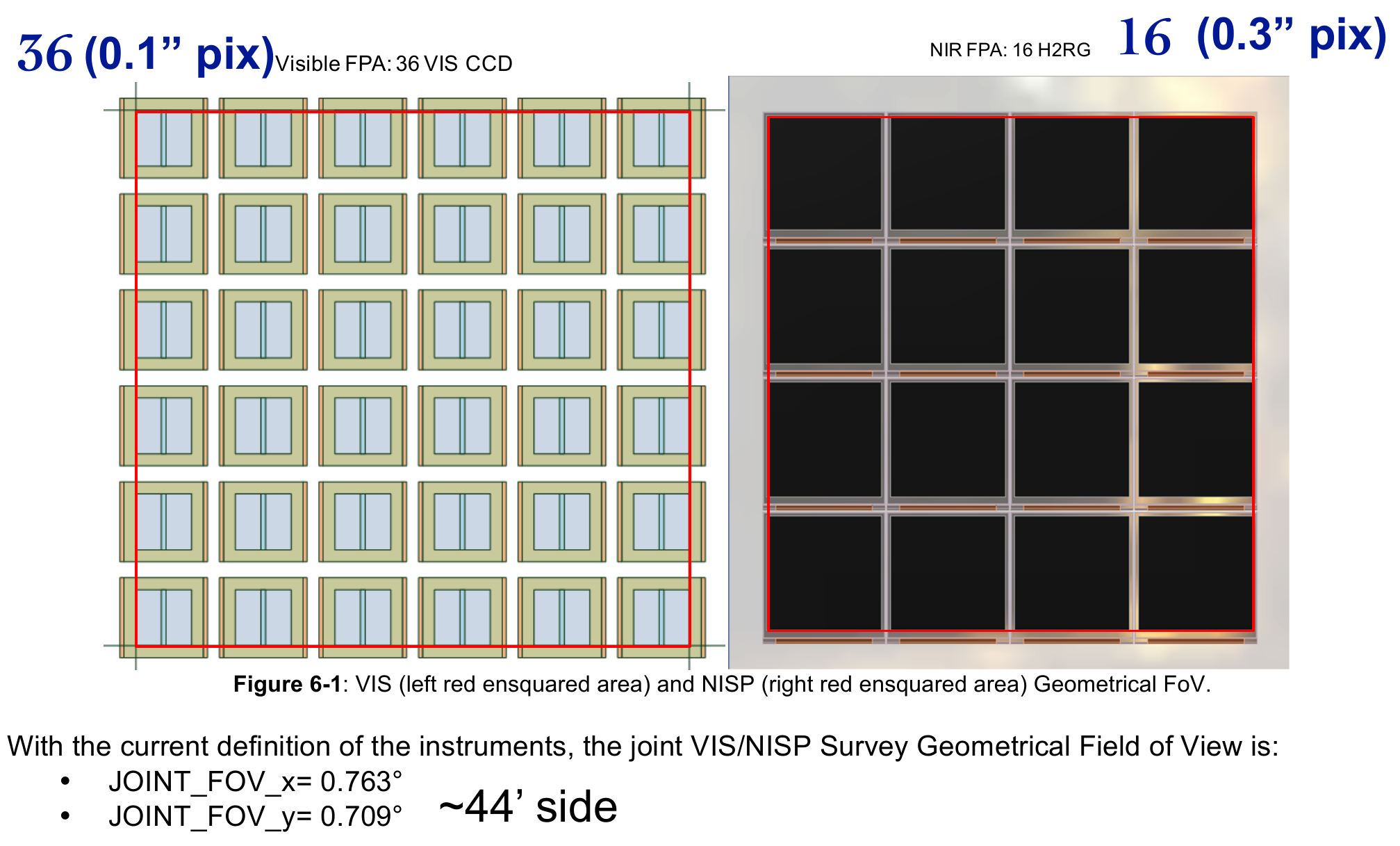} 
\caption{The focal plane of the two instruments VIS and NISP. The joint FoV is $\sim$0.5 sq. degs.}
   \label{fig2}
\end{center}
\vspace*{-0.30 cm}
\end{figure}

\noindent  {\underline{\it Instruments and observations}}.
In Euclid direct contributions to ESA are provided by an international consortium [EC], lead by Y. Mellier,  which comprises most of european nations (as of today $\sim$ 1200 participants from over $\sim$150 institutes) and provides the two instruments VIS and NISP. VIS (\cite{VIS}) is an optical imager well matched to the telescope sampling (see Fig. 3), while NISP (\cite{NISP}) can take either slitless spectra or NIR photometry in the Y, J, H bands. EC is also providing the Ground Segment data analysis, which will be carried out in several data centers. Also NASA contributes to the mission with the NIR arrays and therefore a selected number of US scientists participate as full members to the EC.
%
%
\begin{figure}[h]
\floatbox[{\capbeside\thisfloatsetup{capbesideposition={left,center},capbesidewidth=7cm}}]{figure}[\FBwidth]
{\caption{The telescope will operate in a step and stare fashion. 
Precise thermal stability requires to observe orthogonally to the sun, limiting time visibility of areas at low ecliptic latitudes.}\label{fig:telescope}}
{\includegraphics[width=5cm]{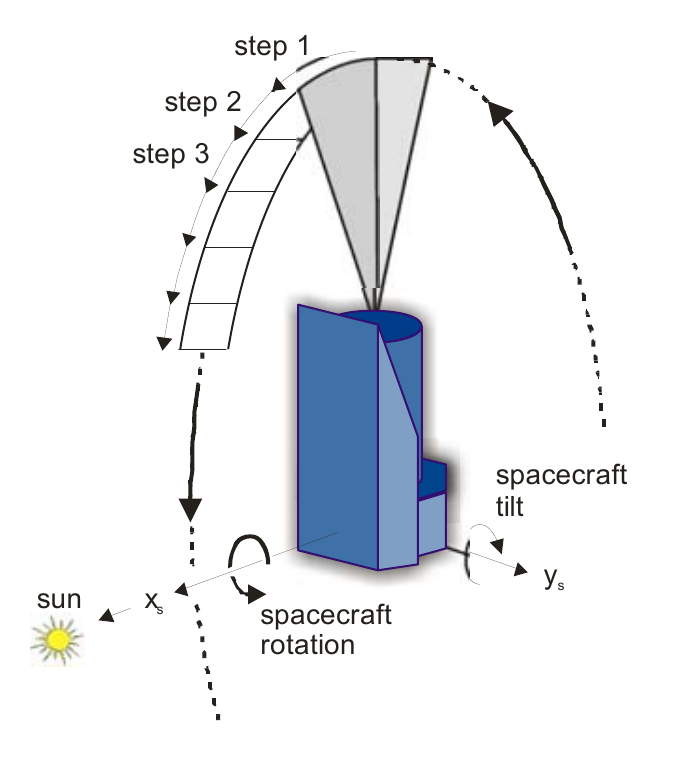}}
\vspace*{-0.50 cm}
\end{figure}
%

The Euclid spacecraft has stringent requirements on operations in order to minimise systematics and maintain the superb optical quality over a large field of view. This in turn translates in the need of observing always orthogonally to the sun and in a step and stare scan strategy (see Fig. 2) which approximately covers $\sim 20$ fields, i.e. $\sim10$ square degrees, per day. The sun-angle constraints imply only few days of visibility for low ecliptic latitudes and that perennial visibility (needed to satisfy specific needs of cadence for calibration purposes) is achieved only over two circles of 3 deg radius centered on the ecliptic poles. This forces to set the position of the deep fields as close as possible to the ecliptic poles. The Southern one cannot be centered because of the location of the Large Magellanic Cloud. Details are given in \cite[Amiaux et al. (2012)]{refsurvey12} and \cite[Tereno et al. (2014)]{thisvolume}.
%
\begin{figure}[h]
\begin{center}
 \includegraphics[width=11cm]{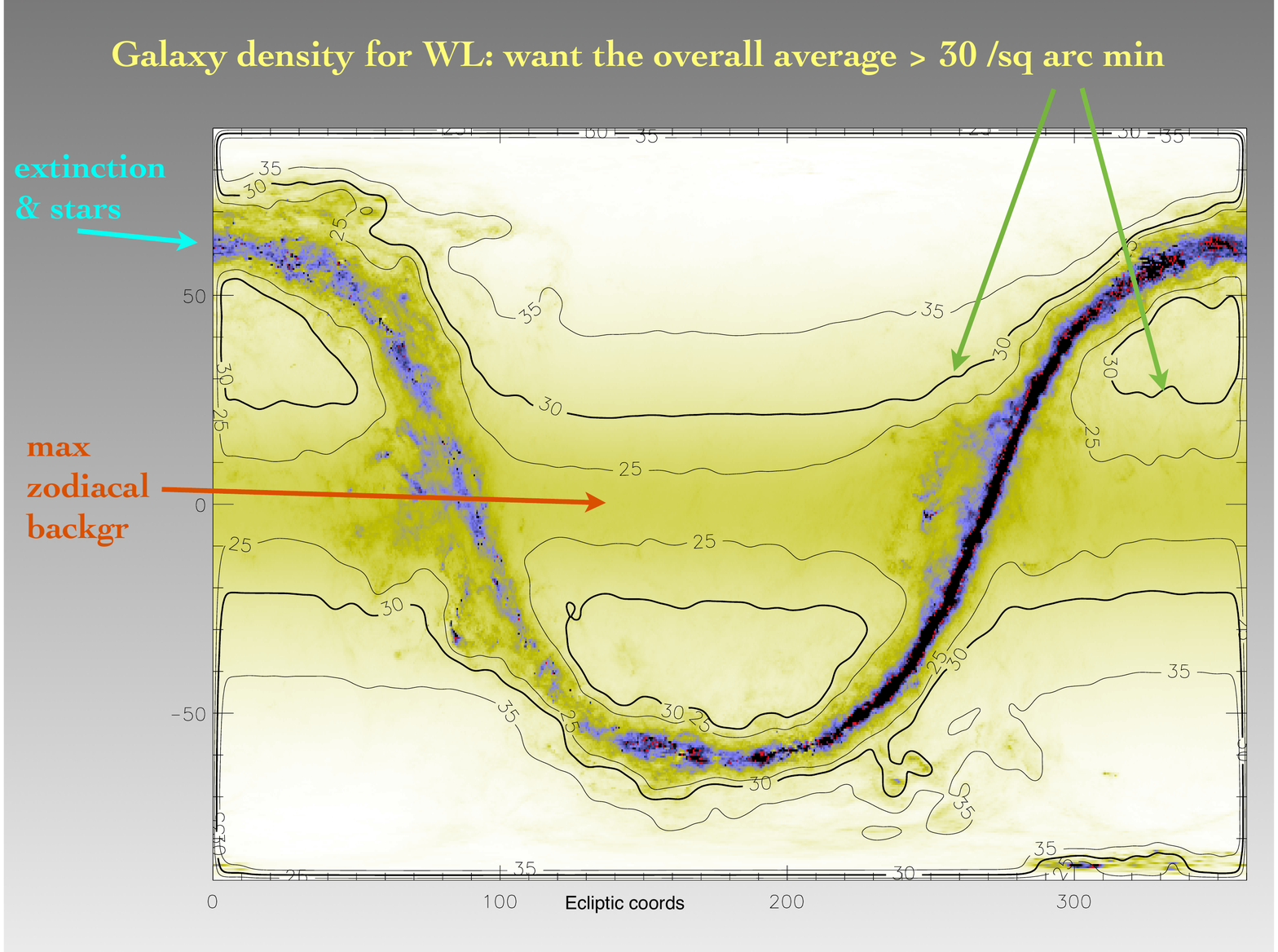} 
 \caption{Euclid expected galaxy density on the sky [counts/arc min$^2$]  g  as a function of average zodiacal background and extinction. These galaxies have S/N$>$10 with a limit of $M_{ab}=24.5$ (wide optical filter) and minimum size  of FWHM[gal]$>$1.25 FWHM[PSF]}
   \label{fig4}
\end{center}
\vspace*{-0.30 cm}
\end{figure}
%

Each field is observed in four identical sequences but for small dithers (\cite{redbook}; \cite{refsurvey12}). In each sequence first the longest exposures for spectra (NISP) and morphology (VIS) are taken at the same time, then these are followed by a sequence of three photometry exposures in Y, J and H band (change in filters would perturb the VIS exposure, so no data are taken during this phase).  The four slitless exposures will be taken with a red grism ($\sim$ 1.25--1.85 $\mu$m) but with three different main orientations for the dispersion direction, to take care of possible line contamination in spectra caused by nearby objects. Also a blue grism ($\sim$ 0.92--1.3 $\mu$m) will be present on board to be used on the deep fields. This will extend the wavelength spectral coverage in a way which is very important for legacy purposes (e.g. line ratios; \cite{redbook}). 
 \vspace*{-0.50 cm}
%
%
%
\noindent \vskip -4truecm
\section{Data models and systematic effects}
The large number of sampled objects (see Fig. 4) and the wide area coverage (see Fig. 5) translate into a good sampling of all scales of interest such that a very good modelling and control of biases is needed for Euclid so to reach its primary science goals. 
\begin{figure}[h]
\vspace*{-1.50 cm}
\floatbox[{\capbeside\thisfloatsetup{capbesideposition={left,center},capbesidewidth=5cm}}]{figure}[\FBwidth]
{ \caption{Window function (binned) coefficients (we plot the customary $\ell (\ell + 1) W_\ell $) obtained via Healpix routines \cite{healpix} for two different sky coverages (\cite{refsurvey12},
\cite{thisvolume}).  }
\label{fig:llcl}}
{\includegraphics[width=9cm]{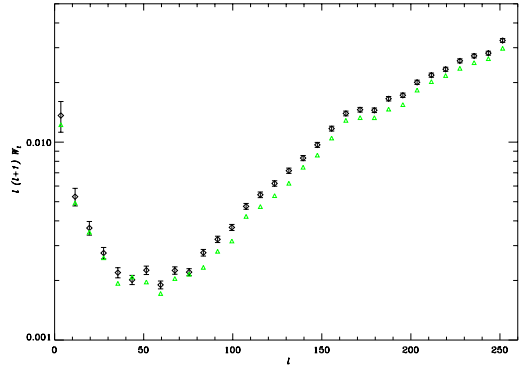}}
\vspace*{-1.0 cm}
\end{figure}
%
Therefore a lot of work continues to be done on these aspects, taking also advantage of experience on real data from ground based surveys or in similar situations (e.g. HST, also slitless). We list some of these efforts which, though often advanced, in part will benefit in the next years also from new or developing techniques in statistical analysis. Important aspects are: 
Shear bias and PSF modelling (\cite{lensinginspace}); 
intrinsic color gradients across galaxy profiles, of particular importance for Euclid because of the wide optical filter
(\cite{colgrad}); the so-called ``noise bias'' (\cite{noisebias}).
Another crucial issue is the Charge Transfer Inefficiency (CTI), which  originates from radiation damage on CCDs. This causes trailing of charges and therefore shape distortions in a complex, non-linear way.  Strenuous efforts went into modelling it in HST, GAIA and Euclid \cite[(Massey et al. 2013]{CTI1}, \cite[2014)]{CTI2}.
Spectral  completeness and purity estimates require extensive and complex 2D simulations of spectra to estimate contamination, extraction and misclassification of emission lines (see \cite{TIPS}) to assess completeness and purity of the spectroscopic sample.
\vspace*{-0.50 cm}
%

\end{document}